\documentclass[12pt]{iopart}
\newcommand{\SC}[2] {\section{\label{sec#1}#2}} 
\newcommand{\FI}[3]{\begin{figure}
\includegraphics{#2} \caption{\label{fig#1}#3} \end{figure}}
\newcommand{\BI}{\begin{itemize}}
\newcommand{\EI}{\end{itemize}}
\newcommand{\BC}{\begin{center}}
\newcommand{\EC}{\end{center}}
\newcommand{\BE}{\begin{equation}}
\newcommand{\EE}{\end{equation}}
\newcommand{\BA}{\begin{eqnarray*}}
\newcommand{\EA}{\end{eqnarray*}}
\newcommand{\PD}[2]{\frac{\partial #1}{\partial #2}}
\newcommand{\TD}[2]{\frac{d #1}{d #2}}
\newcommand{\F}[2]{\frac{#1}{#2}}
\newcommand{\HM}[1]{\mbox{\hspace{#1 cm}}}

\usepackage{graphicx}
\begin{document}

\title{Surface effects on nanowire transport: numerical investigation 
using the Boltzmann equation}

\author{Venkat S. Sundaram and Ari Mizel}

\address{Department of Physics and Materials Research Institute, 
Pennsylvania State University\\ University Park PA 16802.\\ E-mail: ari@phys.psu.edu}

\begin{abstract}
A direct numerical solution of the steady-state Boltzmann
equation in a cylindrical geometry is reported. Finite-size effects are
investigated in large semiconducting nanowires using the relaxation-time
approximation. A nanowire is modelled as a combination of an interior with
local transport parameters identical to those in the bulk, and a finite
surface region across whose width the carrier density decays radially to
zero. The roughness of the surface is incorporated by using lower
relaxation-times there than in the interior.

An argument supported by our numerical results challenges a commonly used
zero-width parametrization of the surface layer \cite{CH}. In the
non-degenerate limit, appropriate for moderately doped semiconductors, a
finite surface width model does produce a positive longitudinal
magneto-conductance, in agreement with existing theory \cite{CH}.  
However, the effect is seen to be quite small (a few per cent) for
realistic values of the wire parameters even at the highest practical
magnetic fields. Physical insights emerging from the results are
discussed.

\end{abstract}

\SC{1}{Introduction}
	The effect of a finite system size on the conductivity of a
material is a subject of considerable physical interest, which has
recently been lent added relevance and importance by rapid developments in
nanowire synthesis and assembly \cite{sy1, sy2, Ag, ec1, sur, InP, CLD,
KK, TM}, electrical characterization and transport measurement methods
\cite{ec1, CL1, CL2, CL3, CL4, H1, H2, H3, CL5, CL6, CL7, CL8, VO}. The
development of nanowires currently represents an important part of
materials and applied physics research. Hierarchial self-assembly
techniques \cite{CL3} envisaged in semiconductor nanowires make them
promising central elements of future integrated electronics.\\

	Reports of basic functional two- and three-terminal semiconductor
nanowire devices including junctions, bipolar transistors and field-effect
transistors are now widely found in the literature. The nano-scale
transport properties of several important semiconductors including Si,
\cite{H1, H2, CL2, ec1}, GaAs \cite{CL7}, GaN\cite{CL6} and Ge \cite{Ge,
Ge2, GaP}), and semi-metals \cite{Se, Sb, Bi, Bi0, Bi1, Bi2, Bi3, Bi4}
have been investigated in detail. (Bi has attracted
great attention because of its unique combination of interesting
properties and its potential for thermoelectric applications \cite{Bi,
Bi0, Bi1, Bi2, Bi3, Bi4, th1, th2}.) Bicrystalline nanowires \cite{KK},
crossed nanowire structures \cite{CL2}, functional networks \cite{CL3,
CL8, H3} including ultra-high-density lattices \cite{H3}, heterostructures
\cite{het} and superlattice devices \cite{CL7} are part of the rapidly
growing body of novel nanowire configurations under development.\\

	A central aspect of theoretical enquiry must be the extent to
which the conductivity of a nanowire differs from that of the bulk
material. At first glance, it seems reasonable to suppose that the
conductivity is smaller in a nanowire because of the addition of surface
scattering, assuming that the band structure does not change drastically.
However, experimental results have yielded conflicting indications on this
point, which is presently not well-understood. Particularly intriguing are
reports of mobility values higher than their bulk counterparts observed in
silicon nanowires \cite{CL5}; while in other cases the mobility has been
deduced to be orders of magnitude lower.\\

	Such challenging theoretical questions, brought into immediate
relevance by the extensive data on electrical transport in nanowires
compiled in the last few years, make a thorough quantitative investigation
valuable at this point. The widespread pursuit of experiments pertinent to
the surface effect on conductivity motivates a generic numerical
description of the finite-size effect allowing both freedom and simplicity
in the incorporation of nanowire characteristics.\\

        For large- and moderate- sized nanowires operated at room temperature,
semi-classical kinetic effects are expected to be as important as quantum
mechanical effects like the modification of band structure. Thus a
semi-classical study is both necessary and desirable in the common regime
where the diameter of the nanowires is much larger than the thermal
wavelength. While a quantum mechanical approach is indispensable to
investigate conductance in very narrow nanowires (radii of a few nm)  
\cite{UZ, ZUNG}, it is neither viable nor suitable for larger nanowires
with radii of the order of 100 nm. For the latter the use of the
semi-classical Boltzmann equation is the appropriate method to
approach a thorough quantitative study of finite-size effects.\\

	These effects have previously been addressed theoretically only
with simplistic assumptions regarding the surface. The analytical results
most widely quoted are due to Chambers \cite{CH}, who used
kinetic-theoretical ideas to calculate the modification of the effective
mean free path due to the presence of a surface and then used this value
to find the conductivity. Essentially the same results were obtained using
the Boltzmann equation with suitable boundary conditions in \cite{DING}.\\

	We present here the results obtained from a general numerical
scheme we have developed to solve the Boltzmann equation in a
cylindrical geometry. Such a direct numerical solution offers the
requisite freedom in incorporating nanowire characteristics and
conditions such as the equilibrium electron density profile, or the
presence of defects and impurities. To retain conceptual simplicity,
we employ the relaxation-time approximation as a first step towards
the systematic modelling and prediction of nanowire conductivity with
a view to relating model parameters to experimental data, and possibly
casting a light on surface characterization \cite{sur, sur1}.  In this
context, comprehensive experimental studies of Bi nanowires are of
direct relevance \cite{Bi, Bi2, Bi3, Bi4}. The introduction of
diameter-controlled synthesis of nanowires \cite{CLD, sur} provides
yet another fruitful context for our study.\\

        We emphasize that surface scattering is just one of many scattering
mechanisms that contribute to the resistivity of a nanowire; other
mechanisms, especially acoustic phonon scattering, can be produce more
dissipation in many circumstances.  In addition, for studies of
specific nanowires with specific surface defects and impurities, the
relaxation time approximation employed here would ideally be replaced
by a more exact approach such as Monte Carlo or quantum mechanical
simulation as mentioned above.  Our intention, however, is to
contribute generic intuition about the behavior of surface scattering
in large wires, even when the specific surface scattering centers are
unknown and even when other resistive effects may be primary.  For
this reason, we use a generic relaxation time approach and map out its
predictions for varied choices of wire characteristics.  Despite its
approximate character, this approach confers useful insight and is
substantially more refined than the often-invoked "specularity
coefficient" model \cite{CH,DING}. \\

	Inter alia, the results yield insights into the general problem of
transport with a nominal relaxation time that varies with spatial
coordinates, which is non-trivial because of the fact that the diffusion
of carriers connects different spatial regions, making their properties
inter-dependent, and thereby introduces a connectivity to the physical
situation. This may be of direct relevance to transport in layered media
such as magnetic multi-layers \cite{MM1, MM2, MM3, MM4}.\\

	 The numerical problem resulting from a finite-difference
representation of the Boltzmann equation with a simple grid is solved
using the conjugate-gradient method, which offers significant
computational efficiency. A comparison with analytical results in limiting
cases confirms the reliability of the scheme.\\

	The paper is organized as follows: In section \ref{sec2} we
describe our computational framework, and the form of the Boltzmann
equation adapted to the problem at hand. Our surface model is described in
section \ref{sec3}. We present our results in the simple case of
zero--magnetic-field in section \ref{sec4}, showing how they conform to
physical expectation, thus validating the numerical scheme used. The
important limit of zero-surface width is considered in section \ref{sec5}.
Section \ref{sec6} is devoted to the longitudinal magneto-conductance
arising due to the surface effect and is followed by a summary (section
\ref{sec7}).

\SC{2}{Boltzmann Equation and Problem Specification}
	In our approach, we consider a cylindrical conductor which has a
finite surface width, so that there is no abrupt change at the boundary
to be dealt with through boundary conditions. In particular, this
includes treating the unperturbed (i.e. when the external fields are zero)
distribution function $f_0$ as a function of both space and momenta. In
order to use the Boltzmann equation with such a distribution function
$f_0$ we introduce an effective internal electric field $\bf E_c(\bf r)$ 
in addition to any external field, to account for this spatial variation 
of $f_0$. In a confined cylindrical system where $f_0$ decays from its interior
magnitude to zero over a finite surface width, the force due to this
internal field is exactly analogous to the constraining force that keeps a
particle within an enclosure with an abrupt boundary, and is therefore
physically expected. With this additional internal field, the Boltzmann
equation \cite{TXT} for the distribution $f({\bf p}, {\bf r})$ of
non-interacting carriers in the relaxation time approximation with a
spatially varying relaxation time is
\BE \nabla f \cdot \F{\bf p}{m} + \nabla_{\bf P}f \cdot q({\bf E} +
{\bf E_c} + \F{\bf p}{m} \times {\bf B}) = -\F{f - f_0}{\tau({\bf r})} \EE
Here {\bf E} and {\bf B} are the external electric field and magnetic
field, and $q$ and $m$ are the charge and mass of each carrier.
It is convenient to introduce the deviation $\phi({\bf p}, {\bf r}) =
f({\bf p}, {\bf r}) - f_0({\bf p}, {\bf r})$ due to the presence
of the external fields; the equation for $\phi$ is
\BE \phi = - \tau({\bf r}) [ \nabla \phi \cdot \F{\bf p}{m}
+ q\nabla_P \phi \cdot ({\bf E + E_C} + \F{\bf p}{m} \times {\bf B})
+ q\nabla_P f_0 \cdot ({\bf E} + \F{\bf p}{m} \times {\bf B})]. \EE 
We consider a long cylinder with no azimuthal or axial inhomogeneity and
confine attention to the case where the external fields are uniform and
parallel to the axis of the wire, defining a natural axis $\hat{z}$ of
reference: ${\bf E} = E \: \hat{z}; \:\:\:{\bf B} = B \:  \hat{z}\:\:$
(see Fig.~\ref{fig1}.)\\

\FI{1}{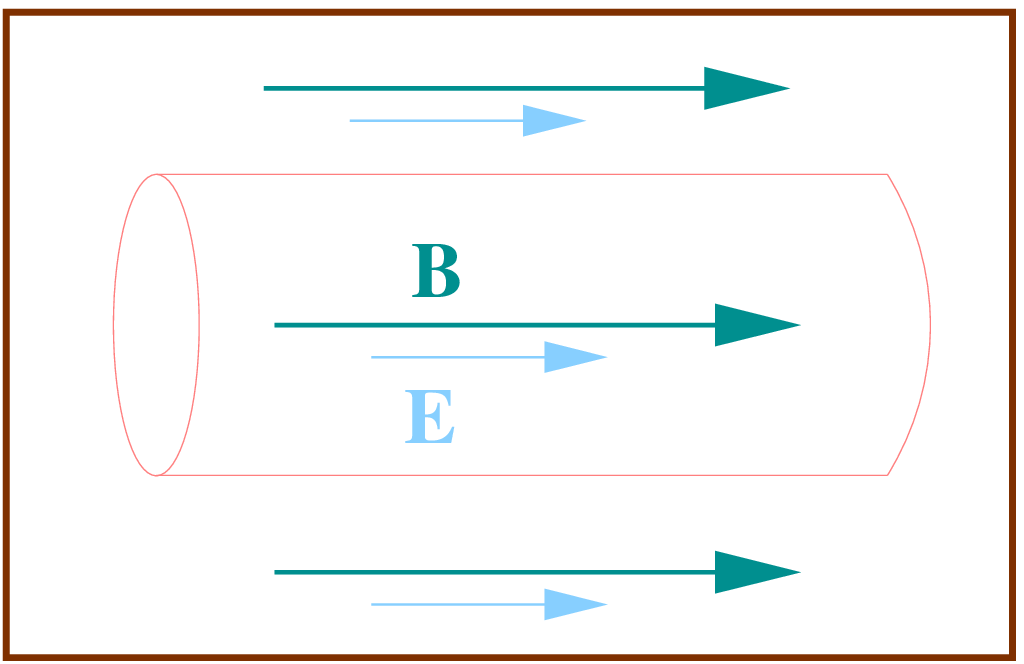}{Model Geometry} Thus the physical problem at hand requires
only one spatial degree of freedom $r = \sqrt{x^2 + y^2}$, although we
have to treat all the momentum components. It is convenient to work with
the local momentum components $p_r = {\bf p}\cdot {\hat r}$ and
$p_{\theta} = {\bf p}\cdot {\hat \theta}$ rather than the canonical
momenta in a cylindrical system, because the canonical momentum conjugate
to $\theta$ is an angular momentum. Also, to preserve form (2) of the
Boltzmann equation, $f$, $f_0$ and $\phi$ are defined as coordinate
densities in the space of $(p_r, p_{\theta}, p_z, r)$ so that $f dp_r
dp_{\theta} dp_z dr$ is the number of carriers in a volume element of this
space. The missing volume element factor $2\pi r$ is absorbed into $f$,
changing the internal field $E_c$, as we shall see. We solve equation (2)
by discretizing on a real space grid of about $20^4$ grid points. The
resulting matrix equation is solved by the method of conjugate gradients.\\

The drift velocity, which is the population average of the component of 
the carrier velocity parallel to the external electric field, is
\BE \langle v_z \rangle = \F{\langle p_z \rangle}{m}
= \F{1}{m} \F{\int d^3p \:dr f({\bf p}, r) p_z}
{\int d^3p\: dr f({\bf p}, r)}
= \F{1}{m} \F{\int d^3p \:dr \phi({\bf p}, r) p_z}
{\int d^3p\: dr f_0({\bf p}, r)} \EE
The effective conductivity $\sigma$ is $q \langle v_z \rangle N / lA$
where $N$ is the total number of carriers in the wire, $l$ is its length 
and $A$ its cross-sectional area.\\

An explicit expression for $E_c$ can be directly deduced from the spatial
variation of $f_0$ for an arbitrary form of the latter, since by
definition of the unperturbed distribution,
\BE \nabla f_0 \cdot \F{\bf p}{m} + \nabla_{\bf P}f_0 \cdot q
{\bf E_c} = 0 \EE
For simplicity, we assume that the spatial dependence of the distribution function is
completely separable from the momentum dependence so that $f_0 = {\it
F}({\bf p}) \xi({\bf r})$ and the internal field ${\bf E_C}$ points only
in the direction of $r$ : ${\bf E_C} = E_C(r) \hat{r}$. The computational
results described below pertain to the case where the
momentum-distribution is Maxwellian:
\BE F({\bf p}) = \exp(-\F{p^2}{2mkT}) \EE
Inserting $f_0 = {\it F}({\bf p}) \xi(r)$ and (5) in (4) we see that the
simplest consistent form of the internal field is
\BE E_c(r) = \F{kT}{q} \F{1}{\xi(r)} \TD{\xi}{r}. \EE

\SC{3}{Surface Model} We use a simple, continuous model to include the
effects of the surface on the conductivity. The fact that there are no
carriers beyond the wire radius is accounted for by taking the volume
density of carriers to decay spatially as a Gaussian with a width $w$
beyond a certain radius $r_0$.  A Gaussian function is chosen because
of its mathematical simplicity and because its qualitative form is
physically reasonable -- we anticipate qualitatively valid physical
conclusions using this form.  This corresponds to a physical field
that is proportional to the difference $r-r_0$ for $r > r_0$ and is
directed towards the centre.  An additional term proportional to $1/r$
in the effective internal field $E_c$ arises on using equation (5) due
to the inclusion of the volume factor $2 \pi r$ in the definition of
$f$, as mentioned in section \ref{sec2}.  \BE f_0 = F({\bf p}) \xi(r)
= {{r \exp(-\F{p^2}{2mkT}) \HM{2.6} r < r_0} \brace
{r\exp(-\F{(r-r_0)^2}{2w^2})\exp(-\F{p^2}{2mkT}) \HM{0.3} r \ge r_0}}
\EE \BE E_c(r) = \F{kT}{qr_0} h(r) = {{\F{kT}{qr} \HM{4.3} r < r_0}
\brace {\F{-kT(r - r_0)}{qw^2} + \F{kT}{qr} \HM{2.3} r \ge r_0}} \EE
\FI{2}{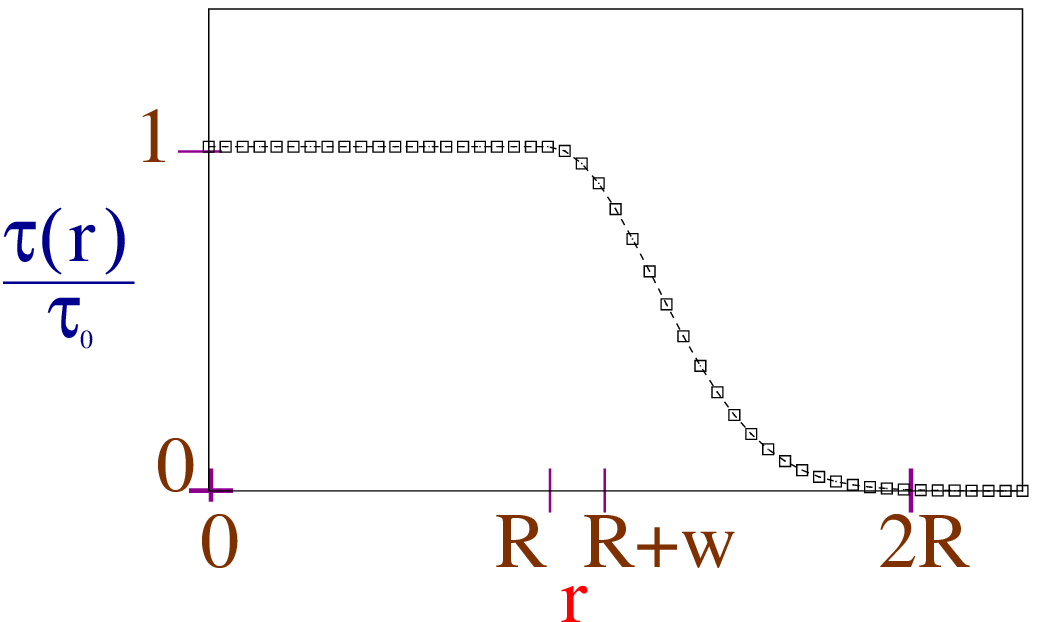}{Radial profile of the relaxation time $\tau(r)$.}
The relative roughness of the surface is described by lower relaxation
times $\tau(r)$ past a radius $R$. The change is also modelled by a
Gaussian fall-off, with a width $w_r$. For simplicity, the radii $R$
and $r_0$ are assumed to coincide, and also the widths $w$ and $w_r$.
The radial profile of the relaxation time is shown in Fig.~\ref{fig2}.
\BE \tau(r) = \tau_0 \chi(r) = {{\tau_0 \HM{4.8} r < R} \brace {\tau_0
\exp(-\F{(r-R)^2}{2w^2}) \HM{2.5} r \ge R}} \EE We take $w$ to be of
the order of $0.1 \times R$.

\SC{4}{Conductivity}
	In the absence of a magnetic field, it is possible to make certain
predictions analytically about the behaviour of the conductivity. We work
in the limit where $E$ is small, so that the drift velocity $\langle v_z
\rangle$ is always small in comparison to the thermal mean velocity
$v_{th}$. If $\tau(r)$ were a constant, $\tau_0$, the drift velocity would
be given by the familiar result \BE \langle v_z \rangle = \F{qE \tau_0}{m}
\EE When $\tau$ acquires a non-trivial radial profile $\tau(r)$, its
overall scale determines to what extent thin cylindrical shells at
different radii affect their neighbours because of the diffusion of
carriers from layer to layer. If this scale is so small that the mean free
path of particles with velocities of the order of $v_{th}$ is much smaller
than the length scale $\zeta$ over which $\tau$ changes significantly 
(i.e. $\tau(r) v_{th} \ll \zeta = w$), the
spatial connectivity is negligible, and one may treat the different layers
separately. In this case, one can define an effective relaxation time
$\tau_{geom}$ which depends only on the geometrical distribution of the
relaxation time weighted by the relative carrier concentration $\xi(r)$.
The use of $\xi(r)$, which is the spatial factor in the unperturbed
distribution function $f_0$, here is consistent with the fact that the
relaxation-time--scale is small. \BE \tau_{geom} = \F{\int dr \xi(r)
\tau(r)}{\int dr \xi(r)} \EE For the model discussed in section
\ref{sec3}, this can be expressed simply in terms of the ratio $\omega = w
/ R$, providing a useful test for the numerical scheme in the limit of
small $\tau_0$. \BE \tau_{geom} = \tau_0 \F{1 + \sqrt{\pi} \omega +
\omega^2} {1 + \sqrt{2 \pi} \omega + 2 \omega^2} \EE

It is therefore useful to consider the essential surface effect as the
departure of the actual drift velocity for the nanowire as a whole from
the value derived by inserting $\tau_{geom}$ in place of $\tau_0$ in (3).
This, as we have just observed, will be perceptible only when $\tau_0
v_{th}$ is comparable to the surface width and significant only
when comparable to the wire radius.\\

Now, it is important to note that the effect of spatially varying $\tau$
in a conductor as a whole is asymmetric in its action, between smooth
(high $\tau$) and rough (low $\tau$) regions. A rough region is much less
affected by the gradient of $\tau(r)$ in its neighbourhood than a smooth
region. This is precisely because spatial connectivity is enhanced
where $\tau(r) v_{th}$ is large as explained previously.
Thus, in a conductor
which has both regions of high $\tau$ and low $\tau$, the latter
separately exhibit geometric (unconnnected)  behaviour, showing no effect
of the presence of the former.\\

But the effect of spatial variation on regions with high $\tau$ is to
decrease mobility there, since carriers there can move enough to sample a
significantly rougher region. Therefore, the effect of spatial variation
on rough regions being negligible, its effect on a conductor as a whole,
too, is a decrease in mobility. This consideration will later appear
prominently in explaining the magneto-conductive effect as well (section
\ref{sec6}).
\FI{4}{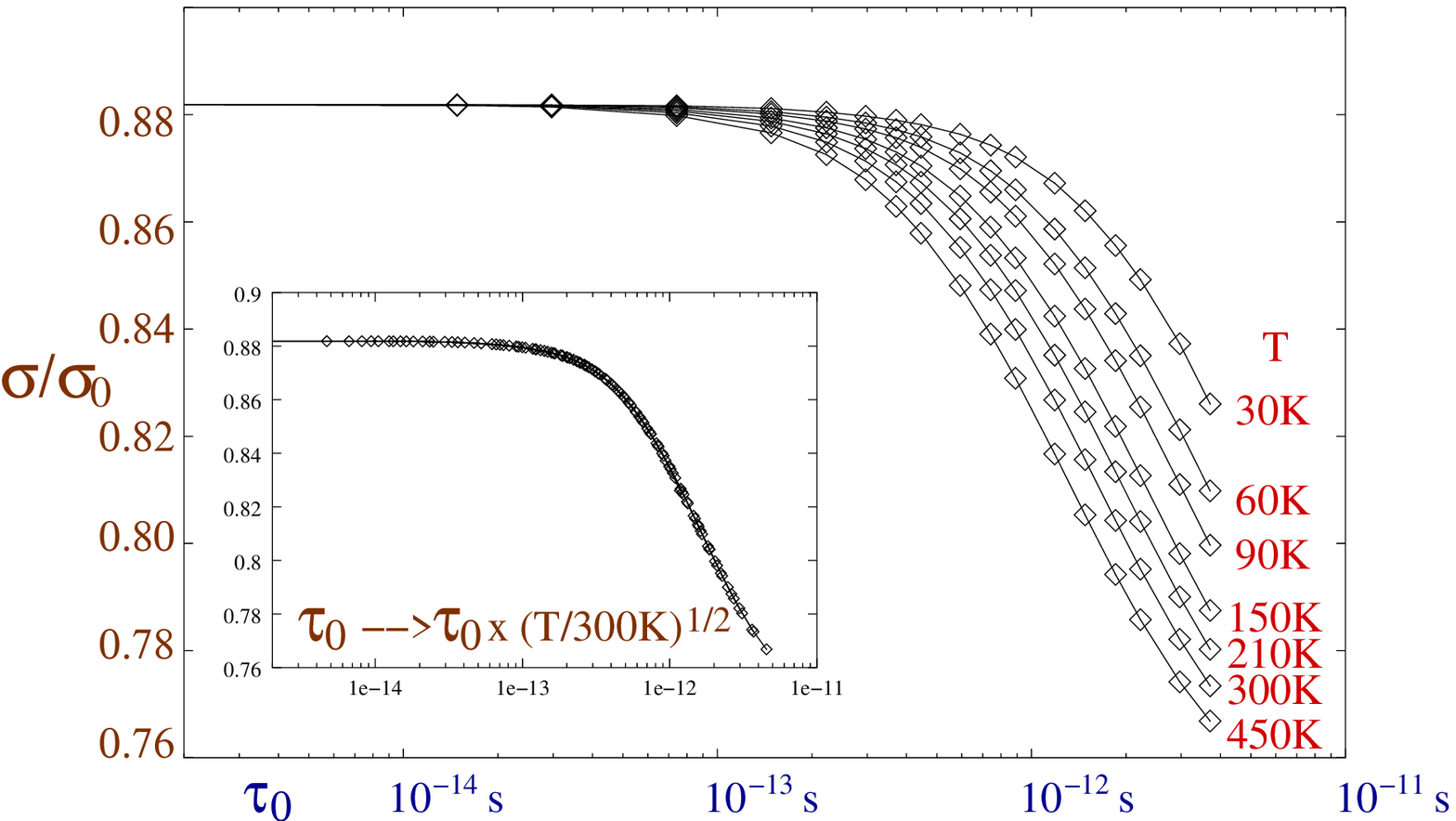}{Relative conductivity as a function of the
relaxation-time scale $\tau_0$ for different temperatures. $\sigma_0$ is
the bulk value of the conductivity corresponding to a constant relaxation
time $\tau_0$. The inset shows the data collapse resulting upon scaling
$\tau_0$ by a factor $\sqrt{T/T_0}$. The fixed parameters are $E = 26 kV/m$,
$R = 200 nm$ and $w = 40 nm$.} Fig.~\ref{fig4} shows the variation of the
conductivity as a function of the relaxation-time--scale for different
temperatures. The ratio of the surface width to the wire radius $\omega$
is fixed at 0.2. As $\tau_0 \rightarrow 0$, the conductivity tends to the
same fraction of the bulk value at all temperatures, which is seen to be
almost exactly equal to the geometric factor 0.8815, the value obtained by
putting $\omega = 0.2$ in equation (12). At higher temperatures, the
departure from this ratio is also higher. This is physically expected
since a higher temperature makes available higher radial velocities to the
carriers in their random motion between collisions, thus increasing the
communication between different layers. Thus we see that even if the
relaxation time were independent of the temperature, the conductivity
would have a temperature-dependence in the low field limit because of the
surface effect. Note that to isolate the surface effect, the complex
temperature variation of relaxation time in a real medium is deliberately
suppressed, though it can be included easily in the computational
scheme.\\

	Further, it is seen in the inset that the data can be collapsed on
to a single curve by using the transformation $\tau_0 \rightarrow \tau_0
\sqrt{T/T_0}$, where $T_0$ is an arbitrary temperature. In other words,
when $B = 0$ the temperature is a reducible parameter in the limit of a
low electric field, which can be accounted for exactly by rescaling $\tau$
in proportion to the corresponding mean thermal velocity. This can be seen
analytically by direct use of the Boltzmann equation, through a
transformation to dimensionless variables. Rewriting eqn. (2)  with the
functional forms in (7), (8) and (9) we have \BE \phi + qE \tau(r)
\PD{\phi}{p_z} + \F{kT}{R} h(r) \tau(r)  \PD{\phi}{p_r} + \tau(r)
\F{p_r}{m} \PD{\phi}{r} = \F{qE \tau(r) p_z}{mkT} e^{-p^2/2mkT}\xi(r) \EE
Now introducing the variables ${\bf u} = {\bf p} / \sqrt{mkT}$ and $s = r
/ R$, we get \BE \phi + \F{qE \tau_0}{\sqrt{mkT}} \chi(s) \PD{\phi}{u_z} +
h(s) \F{\tau_0}{R} \sqrt{\F{kT}{m}} \chi(s) \PD{\phi}{u_r} + \chi(s) u_r
\F{\tau_0}{R} \sqrt{\F{kT}{m}} \PD{\phi}{s} = \F{qE \tau_0 \chi(s) \xi(s)
u_z}{\sqrt{mkT}} e^{-u^2/2} \EE

When E is small, $\phi$ is of the same order of smallness, and therefore
the second term in the left hand side of (14) is negligible. This leaves
us with a linear differential operator containing $\tau_0$ only in the
combination $\tau_0 \sqrt T$ acting on $\phi$ in the left hand side. The
appearance of $\tau_0$ and $T$ in the right hand side (inhomogeneous term)
of course, merely alters the overall scale of $\phi$. Hence the function
$\phi({\bf u}, r)$ may be written as $\F{qE \tau_0}{\sqrt{mkT}} \times
H({\bf u}, s; \tau_0 \sqrt T)$ where H is the solution of (14) with the
constant $\F{qE \tau_0} {\sqrt{mkT}}$ absent in the right hand side.  The
drift velocity is
\BA \langle v_z \rangle &=& \sqrt{\F{kT}{m}} \F{\int d^3u \int ds \: 
u_z \phi({\bf u}, s)}{\int d^3u \int ds \: \xi(s) e^{-u^2/2}}  \\
&=& \F{Eq \tau_0}{m} \F{\int d^3u \int ds \: u_z H({\bf u}, s; \: \tau_0 
\sqrt T)}{\int d^3u \int ds \: \xi(s) e^{-u^2/2}} \\
&=& \langle v_z \rangle_0 \F{\int d^3u \int ds \: u_z H({\bf u}, s; \: 
\tau_0 \sqrt T)}{\int d^3u \int ds \: \xi(s) e^{-u^2/2}} \EA
Thus $\langle v_z \rangle \: / \langle v_z \rangle_0 \:\: (= \sigma / 
\sigma_0)$ depends on $\tau_0$ only through the product $\tau_0 \sqrt 
T$.\\

\FI{3}{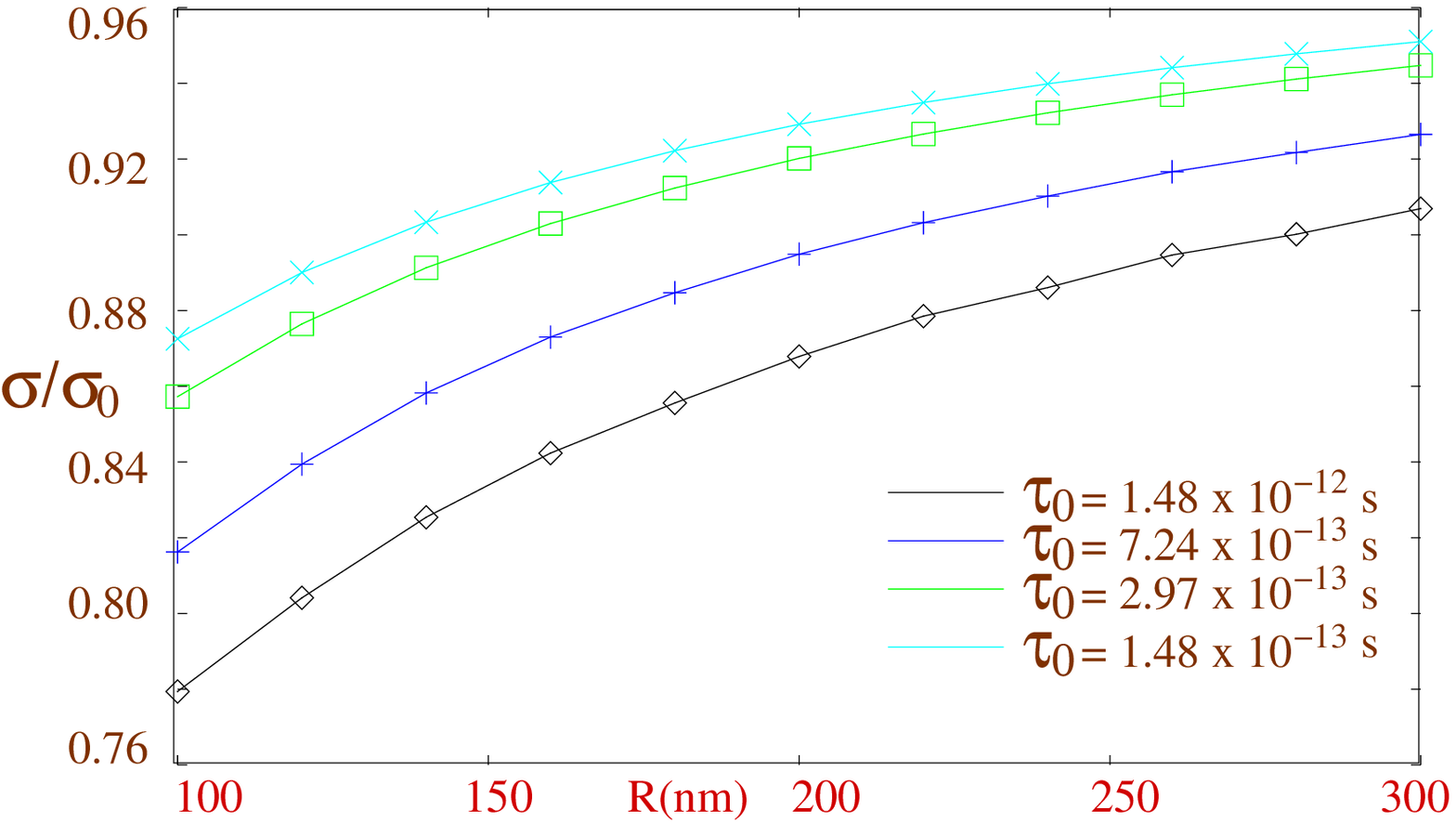}{Relative conductivity as a function of wire radius. Here
$E = 26 kV/m$, $T = 300 K$ and $w = 20 nm$.} Fig.~\ref{fig3} shows the
variation of the conductivity as a fraction of the corresponding bulk
value with wire radius, parametrised by $\tau_0$. As $\tau_0$ decreases,
the results approach geometrically determinable values; further, as
$\omega$ increases, not only does the geometrical factor move further
below $1$ but also the actual conductivity departs more and more from the
geometrical value. Expectably, if the width is fixed, increasing the wire
radius diminishes the strength of the surface effect.\\

	The conformity with physical expectation and analytically known
limits in the results above indicates the reliability of the numerical
scheme and its suitability for other calculations based on the Boltzmann
equation.

\SC{5}{Limit of Zero Surface Width}
	A simple parametrization of the surface in a zero-width surface
model is provided by the "specularity coefficient" $\epsilon$, which is
the non-zero probability of carriers incident on the surface suffering a
scattering event there: they undergo diffuse rather than specular
reflection \cite{ZIM}. This is often encountered in the literature; for 
instance, it has also been used in the analogous context of thermal 
transport by phonons \cite{PH1, PH2, PH3, PH4}.
\FI{5}{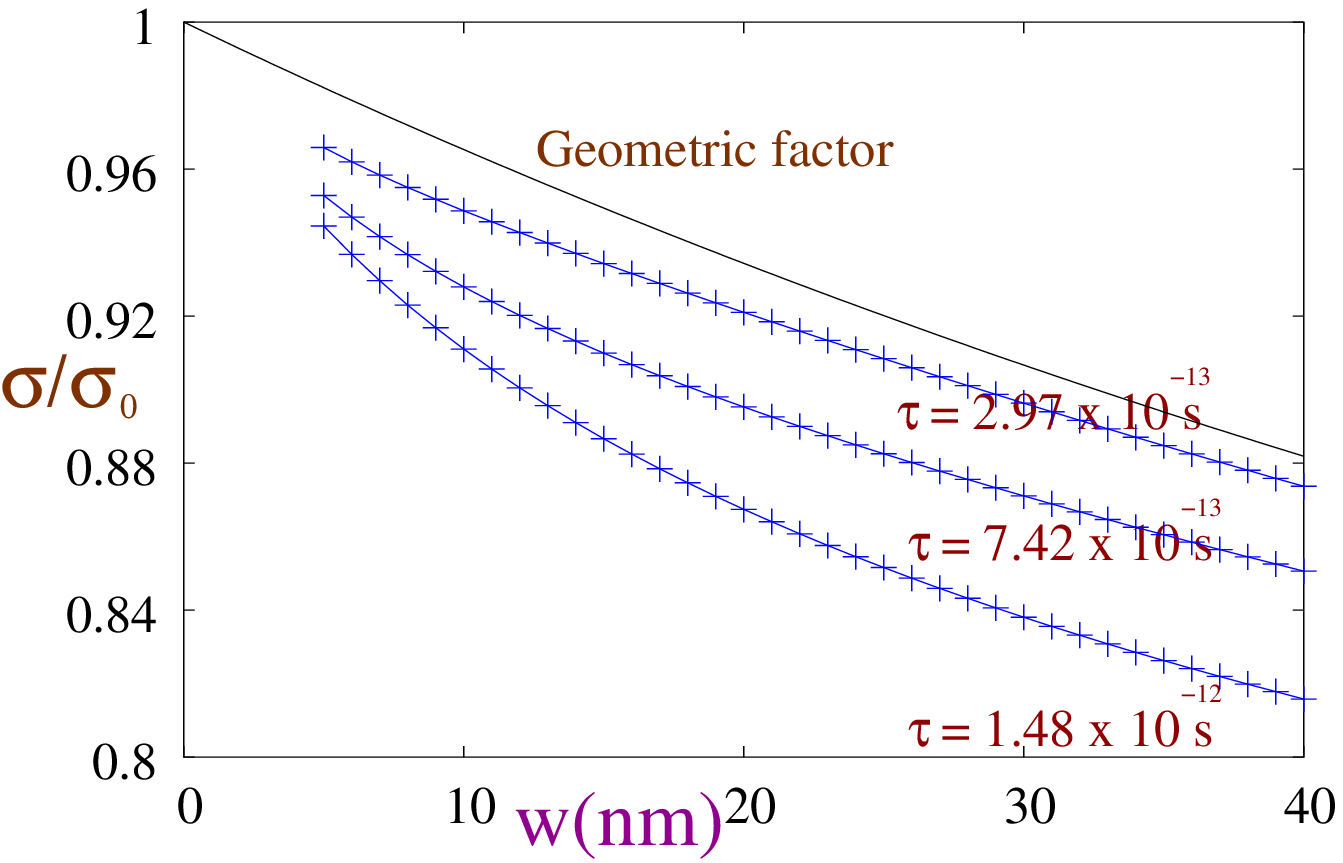}{Variation of
conductivity with surface width for different relaxation-time scales. The
geometric factor is included for reference. Here $E = 26 kV/m$, $T = 300 K$
and $R = 200 nm$.}
We now consider the physically important limit of $w \rightarrow 0$ which
should relate the parameters in our model to the parameter $\epsilon$.
Fig.~\ref{fig5} shows the numerical results obtained by varying the
surface width parameter, along with the graph of the analytical geometric
factor. Significantly, it is seen that even for large $\tau_0$, the
departure from the bulk value seems to go down to a small number, possibly
zero. This is in contrast to \cite{CH} where the magnitude of the surface
effect is unbounded as a function of the relaxation-time scale or mean
free path. It may be noted that the computation becomes increasingly more
expensive, as the surface width is decreased; we believe that the lowest
width used here is low enough to allow us to draw qualitative conclusions
from the results.\\

	We thus see an apparent contradiction between the results obtained
with two different characterisations of the surface, which demands closer
scrutiny. A direct contradiction results only if $\epsilon$ in the abrupt
surface model is considered a free parameter assignable arbitrary values
between 0 and 1; one could interpret the results we have seen as placing
an upper limit on $\epsilon$ not far above zero. But this, on the other
hand, would imply that such a parametrization is not very meaningful.
Indeed, a value of $\epsilon = 1$ is assumed in \cite{CH}.\\

The contrast between the two approaches becomes understandable when one
recognizes the correlation between $\epsilon$ and the fraction of carriers
resident within the surface layer. In a finite width model, this fraction
is always finite; it approaches zero only in the zero width limit, when
$\epsilon \rightarrow 0$. Equivalently, a non-zero value of $\epsilon$ is
only compatible with the existence of a finite fraction of carriers on the
surface. Thus if a finite width model is accepted as more realistic, a
value of $\epsilon$ close to 1 is physically possible only when a large
part of the total number of carriers resides on the surface even in the
absence of a field.\\

There is another way of pointing out the essential difference between an
abrupt surface model and a finite width one. In the former, it is assumed
that a scattering at the surface necessarily returns the carrier into the
bulk of the surface; whereas this is not the case in the latter, once again
bringing to mind the existence of a finite surface population. Finally, in
the former, there is no scale of the relaxation-time at which the surface
effect saturates, whereas, in the finite width model, saturation occurs
where the mean free path exceeds a few times the length scale $\zeta$.
Saturation is physically expected in the latter case -- if in some case,
typical carriers can sample regions with significantly different values of
$\tau$ during a lifetime, any further increase of the lifetime should not
make a big difference.\\

A future investigation of the surface effect resulting from a carrier
distribution involving non-trivial features near the surface rather than
the simple, monotonic decay studied here is expected to provide further
insight. In particular, it may provide an explanation for the apparent
experimental evidence \cite{CHEXP} that $\epsilon$ is close to 1. In any
case, the considerations stated in this section challenge the conception
of the abrupt surface model, which ought to be re-evaluated in the light
of these numerical results.

\SC{6}{Longitudinal Magneto-conductance}
The effect of a longitudinal
magnetic field ($B$) on the conductivity of a large nanowire is of
particular interest because the surface provides a classical kinetic
mechanism for a magneto-resistive effect. In the large field limit, one
would expect the magnetic field to decouple different spatial regions by
confining the carriers kinematically, constraining them to helical motion
between collisions. Thus, when the cyclotron radius is smaller than
$\zeta$, the carriers are effectively localised by the magnetic field, so
that the conductivity is largely determined by the geometric factor. As we
have seen in section \ref{sec4}, the effect of spatial connectivity in a
wire with a rough surface is to lower the conductivity below the geometric
factor. Since a large $B$ field ought to reduce connectivity, the
magneto-conductance clearly ought to be positive. Further, this
magneto-conductance is expected to saturate with $B$ when the cyclotron
radius for carriers with typical momentum values (around the thermal
momentum $p_{th}$) becomes much smaller than $\zeta$. On the other hand,
the magnetic field has an appreciable effect only when the reciprocal of
the cyclotron frequency for carriers with typical momenta is smaller than
the relaxation time; the magnetic field is ineffective if the probability
for a collision before the carrier velocity turns round once is high.\\

\FI{6}{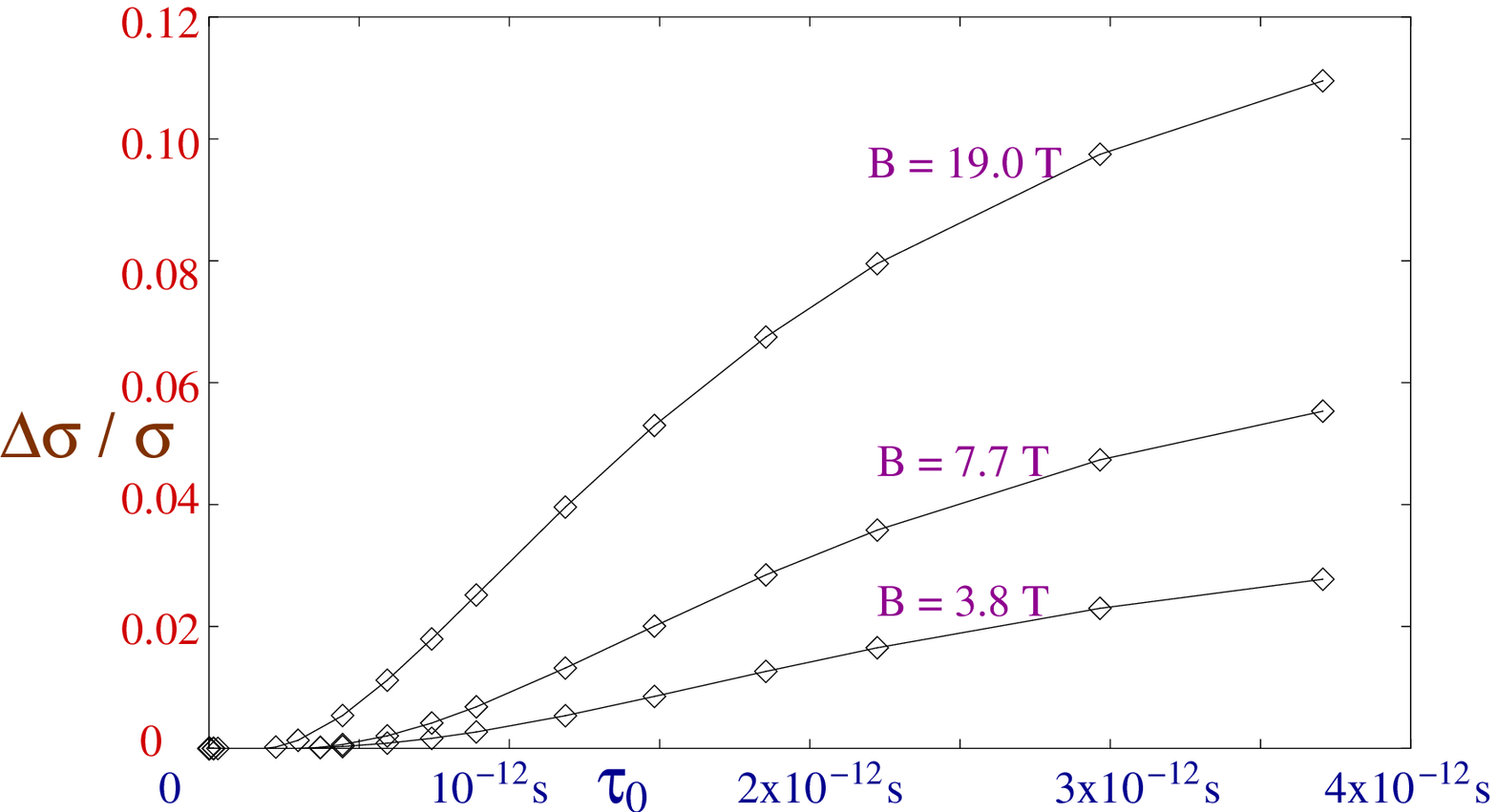}{Fractional longitudinal magneto-conductance as a function
of relaxation-time scale for high magnetic fields. $E = 26kV/m$, $T = 300K$,
$R = 200nm$ and $w = 40nm$.} At intermediate values of $B$, it is not
obvious whether there can be a case of negative magneto-conductance; the
field could conceivably tend to produce a net movement of carriers in some
regions down the gradient of $\tau(r)$. The results presented in
Figs.~\ref{fig6} and \ref{fig7} suggest that this does not occur; the
magneto-conductance is seen to be positive within the accuracy of the
calculation in all cases.\\

	For relaxation times corresponding to the bulk mobility of a
common semi-conducting material like Si, the effect of the magnetic field
is surprisingly small. In particular, the saturation values of $B$ exceed
realistic laboratory values, whereas the magnitude of the fractional
magneto-conductance is still only a few per cent. Again, this is in
contrast to the results in \cite{CH}. Although it is to be noted that the
latter pertain to the degenerate limit (metallic case), the crux of the
difference lies in the surface model, as argued in section \ref{sec5}. In
short, if the surface effect is small, the magneto-conductive effect must
also be limited by the corresponding departure from the geometric
factor.\\

	However, there still remains a qualitative similarity between our
results and the results in \cite{CH}; for instance, we see that the low
$B$ results lend themselves to a good parabolic fit, which is consistent
with \cite{CH}. Further, the arguments we have presented are substantiated
by the fact that \cite{CH} systematically predicts a greater surface
effect than the experimental results used for comparison, a discrepancy
noted there itself. Since the discrepancy is despite the use of best-fit
values, the parametrization itself must be regarded as dubious. It is
important to note, however, that the results in \cite{CH} and our results
are both consistent with recent experimental findings \cite{Bi, Bi0, Bi2}
to the extent that this surface magneto-resistive effect can be
experimentally delineated from intrinsic quantum-mechanical effects.\\

\FI{7}{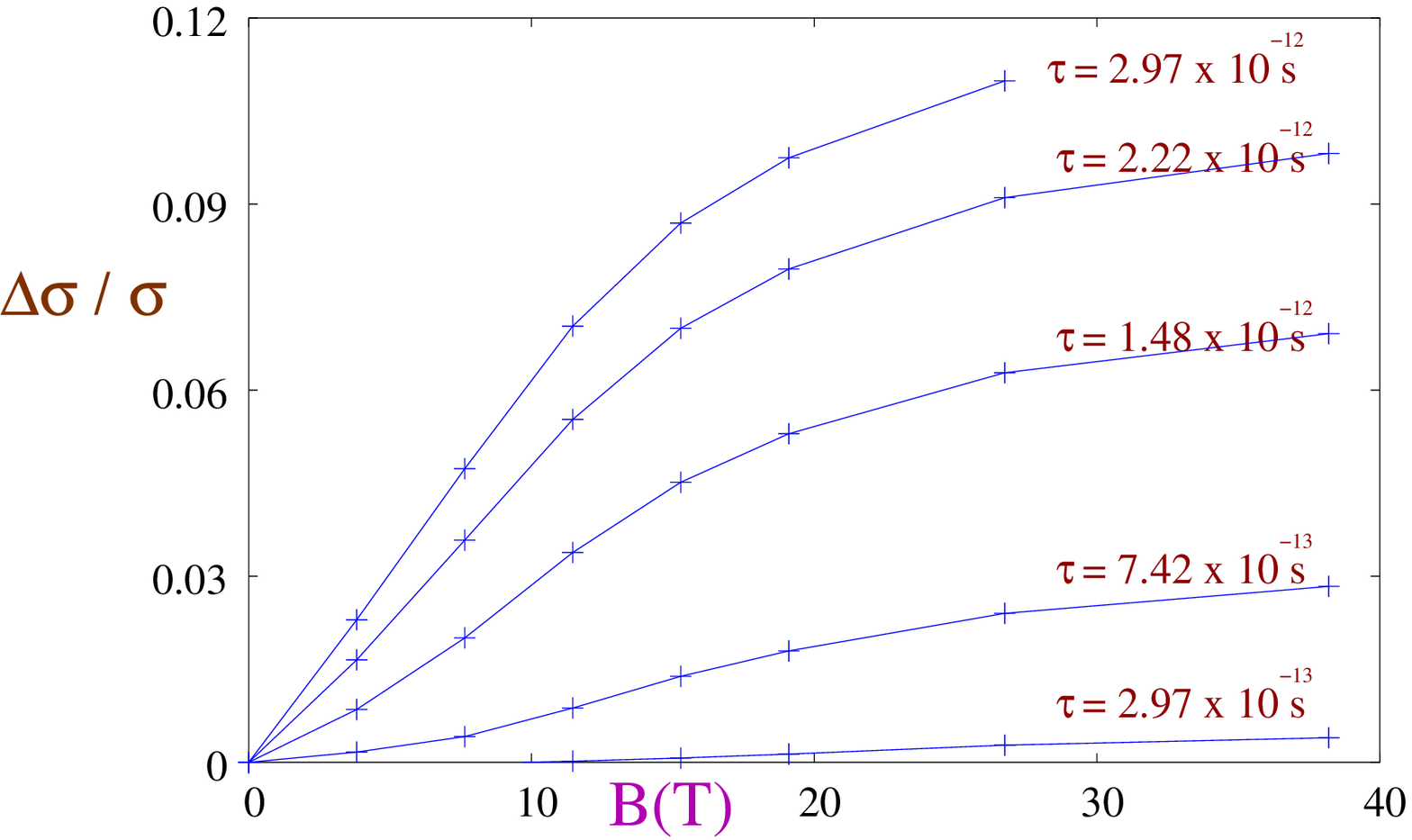}{Variation of fractional magneto-conductance with $B$ 
going up to very large values. Fixed parameters used same as in 
Fig.~\ref{fig6}}
	The intriguing fact that the magnetic field increases conduction
in all cases examined prompts a second look at the general effect of a
spatially varying relaxation time. For one thing, using the same numerical
scheme, we find another curious fact -- the magneto-conductance does not
flip sign when the profile of $\tau(r)$ is inverted, i.e. with a thin
region of high mobility surrounding a core of low mobility, even in the
high $B$ limit. More generally, we could not find a profile of $\tau(r)$
such as to yield a negative magneto-conductance.\\

We attribute the absence of a negative magneto-conductance to the
asymmetry between smooth (high $\tau$) and rough (low $\tau$) regions in
the effect of a spatial variation of $\tau$, discussed in section
\ref{sec4}. As explained there, the surface effect, which is the result of
connectivity in the model at hand, tends to decrease the conductivity
below the geometric value. Thus the introduction of a magnetic field,
whose basic action is to undo the transport connection between different
regions by confining carriers, in reversing the effect of connectivity,
can only increase the conductivity.\\

	One also sees, by the same line of reasoning, that the
surface effect must be small if $\tau(r)$ is assumed to be continuous,
for, the region that contributes most to it is that where both $\tau$ and
its gradient are large; but this region is of the order of the surface
width w, which is expected to be small compared to the wire radius.

\SC{7}{Conclusion}
	A direct solution of the Boltzmann transport equation offers a
powerful approach to transport calculations in large nanowires. We have
presented a simple finite width model in a cylindrical geometry and shown
that despite its simplicity, its scope is significantly greater than that
of an abrupt surface model, from which it exhibits qualitative
differences. This challenges the utility of an abrupt surface model,
especially when parametrized by the specularity coefficient. Our numerical
results show that the classical magneto-conductive effect in large
nanowires of materials like Si is limited to a few per cent even with
magnetic fields beyond the range of a practical laboratory setup.

\ack{The authors thank P. Lammert, T. Mayer, S. Mohney, C. Nisoli, and J. Redwing for
useful discussions.  We acknowledge support under the NSF NIRT program
grant DMR-0103068.
}


\begin{thebibliography}{51}
\bibitem{CH} R. G. Chambers, {\it Proc. Royal Soc. London Ser. A} {\bf 
202}, 378 (1950).
\bibitem{sy1}{
J. D. Holmes, K. P. Johnston, R. C. Doty, and B. A. Korgel, {\it Science} 
{\bf 287}, 1471 (2000).}
\bibitem{sy2}{C. P. Li, X. H. Sun, N. B. Wong, C. S. Lee, and B. K. Teo,
{\it J. Phys. Chem. B} {\bf 106}, 6980 (2002).}
\bibitem{Ag}{Y. Sun, Y. Yin, B. T. Mayers, T. Herricks, and Y. Xia,
{\it Chem. Mater.} {\bf 14}, 4736 (2002).}
\bibitem{ec1}{S. F. Hu, W. Z. Wong, S. S. Liu, Y. C. Wu, C. L. Sung, and 
T. Y. Huang, {\it Solid State Commun.} {\bf 125}, 351 (2003).}
\bibitem{sur}{B. Legrand, D. Deresmes, and D. Stievenard,
{\it J. Vac. Sci. Technol. B} {\bf 20}, 862 (2002).}
\bibitem{InP}{X. F. Duan, Y. Huang, Y. Cui, J. F. Wang, and C. M. Lieber, 
{\it Nature} {\bf 409}, 66 (2001).}
\bibitem{CLD}{Y. Cui, L. J. Lauhon, M. S. Gudiksen, J. F. Wang, and C. M. 
Lieber, {\it Appl. Phys. Lett.}, {\bf 78}, 2214 (2001).}
\bibitem{KK}{H. Carim, K. K. Lew, and J. Redwing, {\it Adv. Mater.} {\bf 
13}, 1489 (2001).}
\bibitem{TM}{J. K. N. Mbindyo, T. E. Mallouk, J. B. Mattzela,
I. Kratochvilova, B. Razavi, T. N. Jackson, T. S. Mayer,
{\it J. Am. Chem. Soc.} {\bf 124}, 4020 (2002).}

\bibitem{CL1}{Y. Cui, Z. Zhong, D. Wang, W. U. Wang, and C. M. Lieber,
{\it Nano. Lett.} {\bf 3}, 149 (2001).}
\bibitem{CL2}{Y. Cui and C. M. Lieber, {\it Science} {\bf 291}, 851 (2001).}
\bibitem{CL3}{Y. Huang, X. Duan, Q. Wei, and C. M. Lieber, 
{\it Science} {\bf 291}, 630 (2001).}
\bibitem{CL4}{Y. Cui, X. Duan, J. Hu, and C. M. Lieber, {\it J. Phys. 
Chem. B}{\bf 104}, 5213 (2000).}

\bibitem{H1}{S. W. Chung, J. Y. Yu, and J. R. Heath,
{\it Appl. Phys. Lett.} {\bf 76}, 2068 (2000).}
\bibitem{H2}{J.Y. Yu, S. W. Chung, and J. R. Heath,
{\it J. Phys. Chem. B} {\bf 104}, 11864 (2000).}
\bibitem{H3}{N. A. Melosh, A. Boukai, F. Diana, B. Gerardot, A. 
Badolato, P. M. Petroff, and J. R. Heath, {\it Science}
{\bf 300}, 112 (2003).}

\bibitem{CL5}{Y. Cui, Z. Zhong, D. Wang, W. U. Wang, and C. M. 
Lieber {\it Nano. Lett.} {\bf 3}, 149 (2003).}
\bibitem{CL6}{Z. Zhong, Fang Qian, D. Wang, and C. M. Lieber {\it Nano. 
Lett.} {\bf 3}, 343 (2003).}
\bibitem{CL7} {M. S. Gudiksen, L. J. Lauhon, J. Wang, D. C. Smith, 
and C. M. Lieber, {\it Nature} {\bf 415}, 617 (2002).}
\bibitem{CL8} {Y. Huang, X. F. Duan. L. J. Lauhon, K. H. Kim
and C. M. Lieber, {\it Science} {\bf 294}, 1313 (2001).}

\bibitem{VO}{J. Muster, G. T. Kim, V. Krstic, J. G. Park, Y. W. Park, 
S. Roth, M. Burghard, {\it Adv. Mater.} {\bf 12}, 420 (2000).}

\bibitem{Ge}{
G. Gu, M. Burghard, G. T. Kim, G. S. Dusberg, P. W. Chiu, V. Krstic,
S. Roth, and W. Q. Han, {\it J. Appl. Phys.} {\bf 90}, 5747 (2001).}
\bibitem{Ge2}{D. W. Wang and H. J. Dai, {\it Angewandte Chemie 
(International Ed.)} {\bf 41}, 4783 (2002).}
\bibitem{GaP}{W. S. Shi, Y. F. Zheng, N. Wang, C. S. Lee, and S. T. Lee,
{\it J. Vac. Sci. Technol. B} {\bf 19}, 1115 (2001).}

\bibitem{Se}{B. Gates, B. Mayers, A. Grossman, and Y. Xia,
{\it Adv. Mater.} {\bf 14}, 1749 (2002).}
\bibitem{Sb}{J. Heremans, C. M. Thrush, Y. M. Lin, S. B. Cronin, and M. S. 
Dresselhaus, {\it Phys. Rev. B} {\bf 63}, 085406 (2001).}
\bibitem{Bi}{Z. B. Zhang, X. Sun, M. S. Dresselhaus, J. Y. Ying, and J. 
Heremans, {\it Phys. Rev. B} {\bf 61}, 4850 (2000).}
\bibitem{Bi0}{Y. Lin, S. B. Cronin, J. Y. Ying, M. S. Dresselhaus, and
J. P. Heremans, {\it Appl. Phys. Lett.} {\bf 76}, 3944 (2000).}
\bibitem{Bi1}{Z. B. Zhang, D. Gekhtman, M. S. Dresselhaus, and J. Y. Ying
{\it Chem. Mater.} {\bf 11}, 1659 (1999).}
\bibitem{Bi2}{K. Liu, C. L. Chien, and P. C. Searson, {\it Phys. Rev. B} 
{\bf 58}, 14681 (1998).}
\bibitem{Bi3}{
T. E. Huber and M. J. Graf, {\it Phys. Rev. B} {\bf 60}, 16880, (1999).}
\bibitem{Bi4}{
T. E. Huber, M. J. Graf, C.A. Foss, and P. Constant {\it J. Mat. Res.} 
{\bf 15}, 1816 (2000).}
\bibitem{th1}{Y. M. Lin, X. Sun, and M. S. Dresselhaus, {\it Phys. Rev. 
B} {\bf 62}, 4610 (2000).}
\bibitem{th2}{X. Sun, Z. Zhang, and M. S. Dresselhaus, {\it Appl. Phys. 
Lett.} {\bf 76}, 3944 (2000).}
\bibitem{het}{M. T. Bjork, B. J. Ohlsson, T. Sass, A. I. Persson, C. 
Thelander, M. H. Magnusson, K. Deppert, L. R. Wallenberg, and L. 
Samuelson, {\it Appl. Phys. Lett.} {\bf 80}, 1058 (2002).}

\bibitem{UZ}{U. Landman, R. N. Barnett, A. G. Scherbakov, and P. Avouris,
{\it Phys. Rev. Let.} {\bf 85}, 1958 (2000).}
\bibitem{ZUNG}{C. Y. Yeh, S. B. Zhang, and A. Zunger, {\it Phys. Rev. B} {\bf 50},
14405 (1994).}
\bibitem{DING} R. B. Dingle, {\it Proc. Royal Soc. London Ser. A} {\bf 
201}, 545 (1950).

\bibitem{sur1}{C. Cadet, D. Deresmes D. Vuillaume, and D. Stievenard,
{\it Appl. Phys. Lett.} {\bf 64}, 2827 (1994).}

\bibitem{MM1}{
A. Barthelemy and A. Fert, {\it Phys. Rev. B} {\bf 43}, 13124 
(1991).}
\bibitem{MM2}{F. Erler, P. Zahn, and I. Mertig, {\it Phys. Rev. B} {\bf 64}, 094408 (2001).}
\bibitem{MM3}{
L. A. Michez, B. J. Hickey, S. Shatz, and N. Wiser {\it Phys. Rev. B}
{\bf 67}, 092402 (2003).}
\bibitem{MM4}{P. B. Visscher, {\it Phys. Rev. B} {\bf 49}, 3907 (1994).}

\bibitem{TXT}{
A. Haug, {\it Theoretical Solid State Physics 46} (Ed. D. Ter 
Haar) Vol. 2 pp. 63-69, Pergamon Press, Oxford (1972).}
\bibitem{ZIM}{J. M. Ziman, {\it Electrons and Phonons}, pp. 456,
Clarendon Press, Oxford (1960).}
\bibitem{PH1}{
J. Zou and A. Balandin {\it J. Appl. Phys} {\bf 89}, 2932 (2001).}
\bibitem{PH2}{
S. G. Walkauskas, D. A. Broido, K. Kempa, and T. L. Reinecke,
{\it J. Appl. Phys.} {\bf 85}, 2579 (1999).}
\bibitem{PH3}{X. Lu and J. H. Chu, {\it Euro. Phys. J. B}
{\bf 26}, 375 (2002).}
\bibitem{PH4}{S. G. Volz and G. Chen, {\it Appl. Phys. Lett.}
{\bf 75}, 2056 (1999).}
\bibitem{CHEXP}{R. G. Chambers, {\it Nature} {\bf 165}, 239 (1950).}
\end{thebibliography}
\end{document}